# Scalable Synthesis of Large-Area $WS_2$ Thin Films from Tungsten Precursors by Thermal CVD and Their Application in Schottky-Barrier Solar Cells

Jisun Kim[1], Yoosuk Kim[2], Seung-Ho Park[1], Yong Hun Ko[1], and Chong-Yun Park[1*]

[1] Department of Physics, Sungkyunkwan University, Suwon 440-746, Republic of Korea

[2] Department of Physics, Saint Louis University, St. Louis, Missouri 63103, United States

*Corresponding author: cypark@skku.edu

## Abstract

Two-dimensional tungsten disulfide ($WS_2$) is a promising semiconductor for next-generation optoelectronic and photovoltaic devices, but scalable routes to uniform, large-area films remain challenging. In this study, a systematic thermal chemical vapor deposition (T-CVD) strategy is presented to synthesize centimeter-scale $WS_2$ thin films by sulfurizing tungsten (W) precursors in a controlled sulfur vapor environment. High-purity sputtered W thin films on $SiO_2$/Si and W foils were sulfurized at temperatures between 400 and 1000 °C, with Raman spectroscopy identifying 800 °C as the optimal growth temperature. Under these conditions, the films exhibit the characteristic $E^1_{2g}$ (349.7 $cm^{-1}$) and $A_{1g}$ (416.8 $cm^{-1}$) modes with narrow full-width at half-maximum values, indicative of high crystallinity and controlled thickness. Optical microscopy, scanning electron microscopy, atomic force microscopy, transmission electron microscopy, and X-ray photoelectron spectroscopy collectively validate the creation of stoichiometric, layered $WS_2$ with significantly enhanced uniformity and diminished roughness when utilizing sputtered W thin films in contrast to W foils. Leveraging this optimized process, $WS_2$ films grown on W foils were transferred onto target substrates, including indium tin oxide (ITO)-coated glass, using a PMMA-assisted wet-transfer method that preserves structural integrity over large areas. Schottky-barrier solar cells with an Au/$WS_2$/ITO architecture fabricated from these films deliver a short-circuit current density of 7.91 mA·$cm^{-2}$, an open-circuit voltage of 0.495 V, and a power conversion efficiency of 1.45 %. These results demonstrate that sulfurization of W thin films and foils via T-CVD provides a scalable, substrate-compatible platform for integrating $WS_2$ into practical optoelectronic and low-cost photovoltaic technologies.

# 1. Introduction

Two-dimensional (2D) materials have attracted extensive interest owing to their reduced dimensionality, tunable electronic structure, and emergent quantum phenomena that differ markedly from those of bulk crystals [1–5]. Since the first demonstrations of graphene's exceptional carrier mobility and Dirac fermion behavior [1,6], the research focus has rapidly expanded to a broader family of 2D materials, including transition metal dichalcogenides (TMDs), hexagonal boron nitride, and various van der Waals heterostructures [2,7–11]. Among these, semiconducting group-VI TMDs such as molybdenum disulfide ($MoS_2$), tungsten disulfide ($WS_2$), and related compounds offer band gaps in the visible to near-infrared range and exhibit strong spin-orbit coupling and valley-selective physics, making them promising for electronic, optoelectronic, and quantum devices [7–13]. Moreover, their electronic structure can be continuously tuned via external perturbations such as strain, enabling systematic modulation of the band gap and optical response in atomically thin TMDs [14].

$WS_2$, in particular, combines a suitable band gap with significant absorption coefficients and strong excitonic effects, enabling efficient light–matter interaction in atomically thin layers [15,16]. Its mechanical flexibility, relatively low density, and tolerance to harsh environments further motivate its use in flexible, wearable, and aerospace-relevant systems [17–21]. Recent studies have proposed $WS_2$-based ultrathin solar cells for space applications, where radiation hardness and specific power are critical design metrics [19]. In this context, achieving large-area, high-quality $WS_2$ films with controllable thickness and morphology is essential for realizing practical photovoltaic and optoelectronic devices that can operate reliably under mechanical deformation and in extreme environments.

Despite substantial progress in exfoliated and chemical vapor deposition (CVD)-grown TMDs [11,22–25], several challenges remain for scalable WS2 thin films directly compatible with device manufacturing. Conventional CVD approaches typically rely on solid or liquid metal-chalcogenide precursors, often yielding triangular monolayer islands or discontinuous domains whose lateral size and layer number are sensitive to precursor composition, temperature, and gas environment [22–27]. Moreover, the structural and chemical evolution of $WS_2$ synthesized via sulfurization of metallic W or tungsten suboxides remains poorly understood, particularly when comparing sputtered W films with bulk W foils. While previous studies have demonstrated $WS_2$ growth on pre-reduced tungsten suboxide substrates [28], a systematic comparison of the resulting film uniformity, roughness, and crystallinity as a function of the initial W form (thin film vs. foil) remains lacking. Such understanding is crucial for optimizing layer control and minimizing defects that degrade carrier transport and recombination in devices [29–31].

On the device side, TMD-based optoelectronics and photovoltaics have progressed from proof-of-concept phototransistors and sensors to more complex architectures, including p-n diodes, van der Waals heterostructures, and Schottky-barrier devices [15,29,32–38]. In particular, Schottky-barrier solar cells employing metal/TMD/transparent-conductor stacks offer a structurally simple route to large-area, semitransparent, and potentially flexible power generators [19,36,38]. By appropriately selecting the metal work function and engineering the metal-semiconductor interface, it is possible to tailor the Schottky barrier height, suppress Fermi-level pinning, and enhance carrier extraction [34–37]. At the same time, the performance of such devices remains highly sensitive to TMD film quality, contact resistance, and nonradiative recombination induced by defects or interfaces [31,38–42]. For $WS_2$ specifically, most photovoltaic demonstrations have relied either on mechanically exfoliated flakes, which are limited in area and difficult to scale [15,43], or on CVD-grown films whose growth conditions and transfer routes may introduce inhomogeneity or damage [24,25,39,40].

These considerations highlight a clear need for a scalable $WS_2$ synthesis route that (i) produces large-area thin films with controllable thickness, high crystallinity, and low roughness; (ii) allows direct comparison between different W precursors (sputtered films vs foils) to elucidate growth mechanisms and morphology control; and (iii) integrates seamlessly with transfer and device fabrication processes for Schottky-barrier solar cells. In this work, we address these challenges by employing a custom-built thermal CVD (T-CVD) system to sulfurize both sputtered W thin films and W foils over a wide temperature range (400-1000 °C). Raman spectroscopy, combined with full-width-at-half-maximum analysis of the $E^1_{2g}$ and $A_{1g}$ modes, identifies 800 °C as the optimal synthesis temperature. In contrast, X-ray photoelectron spectroscopy (XPS) confirms the conversion from tungsten oxide to $WS_2$ [26–28]. Comprehensive scanning electron microscopy (SEM), atomic force microscopy (AFM), and transmission electron microscopy (TEM) characterizations reveal that sputtered-W-derived WS2 forms uniform, wafer-scale films with low surface roughness and well-defined layered crystallinity. In contrast, foil-derived $WS_2$ exhibits thicker, island-like morphologies with significant height variations.

Building on these materials advances, we implement a PMMA-assisted wet-transfer process to integrate large-area $WS_2$ films onto indium tin oxide (ITO)-coated glass and fabricate Au/$WS_2$/ITO Schottky-barrier solar cells. The resulting devices achieve a power conversion efficiency (PCE) of 1.45 %, with performance metrics that compare favorably with earlier $WS_2$- based photovoltaic reports using exfoliated or less-optimized CVD films [15,19,36,39,40,43]. Our results establish sputtered-W-derived WS$_2$ thin films as a scalable platform for WS$_2$ photovoltaics and lay the groundwork for their deployment in low-cost, large-area optoelectronic and energy-harvesting applications, including mechanically flexible and aerospace-relevant systems [19,21,38].

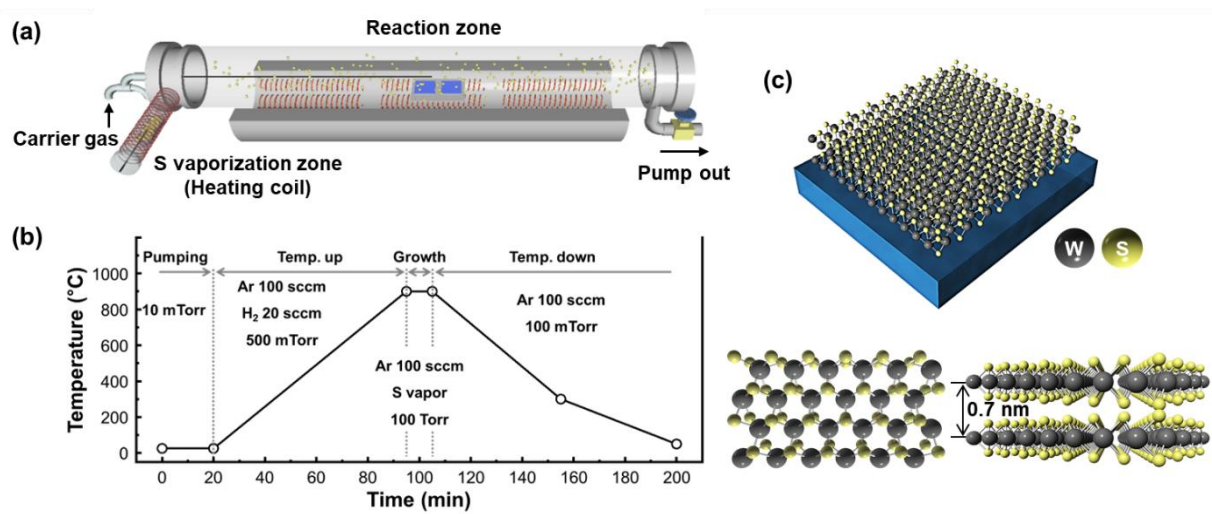

**Figure 1**. **(a)** Schematic representation of the thermal CVD (T-CVD) apparatus employed for developing WS$_2$ thin films. The S vaporization zone is kept at a reduced temperature upstream, while the substrate is positioned in the elevated-temperature reaction zone. The carrier gases (Ar and H$_2$) enter from the left, conveying vaporized S to the substrate; a downstream vacuum pump regulates both flow and pressure. **(b)** Simplified process schematic illustrating the standard growth parameters. During sulfurization, the furnace is maintained at a total pressure of ~100 Torr with an Ar flow rate of 100 sccm and a fixed S source temperature, so that S vapor is produced upstream and transported to the reaction zone, where WS$_2$ is formed on the substrate. **(c)** Atomic configuration of stacked WS$_2$. W atoms (dark gray spheres) constitute a hexagonal plane interposed between two S atom planes (light yellow spheres). The two-dimensional stacking produces van der Waals gaps between neighboring layers, facilitating the formation of ultrathin films.

## 2. Experimental details

### 2.1 Synthesis of WS$_2$ Thin Films

WS$_2$ thin films were synthesized using a custom-built thermal CVD (T-CVD) system, employing W foils (35 μm thick, 99.95 % purity) and sputtered W films on various substrates. **Figure 1(a)** presents a schematic of the T-CVD setup used in this study. Before synthesis, the W substrates were positioned centrally within the T-CVD chamber, as illustrated in **Figure 1(a)**. A ceramic boat containing 50 mg of sulfur (S) powder with 99.9 % purity was placed upstream and heated to 200 °C to ensure consistent vaporization of sulfur at the designated synthesis temperature.

The T-CVD chamber was initially evacuated to a base pressure of 10 mTorr to eliminate residual contaminants. Subsequently, argon (Ar, 100 sccm) and hydrogen ($H_2$, 20 sccm) gases were introduced, and the temperature was increased to the target synthesis temperature, resulting in a chamber pressure of 500 mTorr. Upon reaching the desired temperature, S vapor was introduced, and the chamber pressure was precisely controlled at 100 Torr using a needle valve, providing a stable reaction environment for sulfurization. **Figure 1(b)** plots the variations in pressure, temperature, and gas flow rates throughout the synthesis process.

To determine the optimal conditions for growing high-quality $WS_2$ films, W substrates were sulfurized at temperatures ranging from 400 to 1000 °C. After sulfurization, the furnace was naturally cooled to room temperature under a continuous Ar flow, preserving film integrity. **Figure 1(c)** shows the synthesized $WS_2$ layer exhibiting a characteristic hexagonal structure in which each W plane is sandwiched between two S planes, forming a layered configuration with an interlayer spacing of approximately 0.7 nm.

## 2.2 Measurements and Characterization

The synthesized $WS_2$ thin films were systematically characterized to investigate their structural, morphological, and chemical properties. Surface topography and film continuity were evaluated using scanning electron microscopy (SEM; JEOL JSM-6700F), which enabled the identification of potential defects, such as pinholes or cracks, that could negatively impact device performance. Atomic force microscopy (AFM) measurements accurately assessed film thickness and surface roughness, confirming uniform layer formation across extensive sample areas.

High-resolution transmission electron microscopy (HR-TEM; JEOL JEM-2100F) was utilized to examine the atomic-scale structure and crystallinity of the $WS_2$ films, clearly revealing lattice fringes indicative of the layered arrangement of W and S atoms. Raman spectroscopy (Renishaw, RM1000-Invia) with a 532 nm excitation laser verified the presence of characteristic $WS_2$ vibrational modes, specifically the $E^1_{2g}$ (349.7 cm$^{-1}$) and $A_{1g}$ (416.8 cm$^{-1}$) peaks, confirming phase purity and providing insight into crystallite quality. Complementing Raman spectroscopy, analyses of lateral domain sizes and overall film uniformity were used to ensure consistent device fabrication quality.

Comprehensive chemical analysis was performed using X-ray photoelectron spectroscopy (XPS; AXIS Ultra DLD) equipped with monochromatic Al Kα radiation (hv = 1486.6 eV). All binding energies were calibrated relative to the C 1s peak of adventitious carbon at 284.8 eV. XPS survey scans quantitatively evaluated the $WS_2$ stoichiometry. At the same time, detailed deconvolution of the W 4f and S 2p core-level spectra using CASA XPS software further validated the chemical states of W and S within the thin films.

## 3. Results and discussion

**Figure 2** shows the Raman spectroscopy analysis, which reveals how growth temperature and W precursor type affect the structural quality and thickness of the synthesized $WS_2$ films. **Figure 2(a)** shows representative Raman spectra of $WS_2$ obtained at sulfurization temperatures between 500 and 900 °C. In all cases, two prominent peaks are observed at ~349.7 $cm^{-1}$ and ~416.8 $cm^{-1}$, corresponding to the in-plane $E^1_{2g}$ and out-of-plane $A_{1g}$ vibrational modes of $WS_2$, respectively [26]. At lower (400 °C) and higher (1000 °C) growth temperatures (not shown), no well-defined $WS_2$ modes were detected, indicating that crystalline $WS_2$ does not form under such conditions. This behavior can be attributed to insufficient thermal energy for crystallization at 400 °C, and possible decomposition or sublimation of $WS_2$ at 1000 °C.

The impact of growth temperature on crystalline quality is further quantified in **Figure 2(b)**, which plots the full width at half maximum (FWHM) of the $A_{1g}$ (red circles) and $E^1_{2g}$ (blue squares) modes. A narrower FWHM generally reflects higher crystallinity and fewer structural defects. Both Raman modes exhibit their minimum FWHM at 800 °C, identifying this temperature as the optimum synthesis condition. The schematic insets in **Figure 2(b)** illustrate the atomic vibrations associated with the $E^1_{2g}$ and $A_{1g}$ modes, emphasizing their sensitivity to bonding environment and disorder.

**Figures 2(c) - (e)** present Raman maps of the $A_{1g}$ - $E^1_{2g}$ peak separation, which is commonly used as an indicator of $WS_2$ thickness [26,27]. The color scale ranges from 65 to 70 $cm^{-1}$. For $WS_2$ derived from a 15 nm W film (**Figure 2(c)**), the map is nearly uniform and predominantly red, indicating a relatively consistent and thinner $WS_2$ layer across the probed area. When the W precursor thickness is increased to 20 nm (**Figure 2(d)**), the map becomes predominantly blue, corresponding to a larger $A_{1g}$ - $E^1_{2g}$ separation and thus a thicker WS2 film. These observations are consistent with earlier reports showing that the $A_{1g}$ mode blueshifts and the $E^1_{2g}$ mode redshifts with increasing layer number, leading to a monotonic increase in the peak

separation [26]. In contrast, the WS$_2$ film synthesized from W foil (**Figure 2(e)**) exhibits pronounced spatial variation, with multiple colors present in the map, reflecting a broad distribution of local thicknesses and non-uniform growth.

The single-point Raman spectra in **Figure 2(f)** directly compare WS2 grown from W foil (top, blue curve) and from sputtered W film (bottom, red curve). The $A_{1g}$ - $E^1_{2g}$ separation is clearly larger for the foil-derived sample, confirming that it is significantly thicker than the film-derived WS$_2$. Combined, the temperature-dependent spectra, FWHM analysis, and Raman mapping demonstrate that sulfurization at 800 °C with a sputtered W thin-film precursor yields WS$_2$ with superior crystallinity, uniformity, and controllable thickness, whereas foil-based growth yields thicker, highly non-uniform layers.

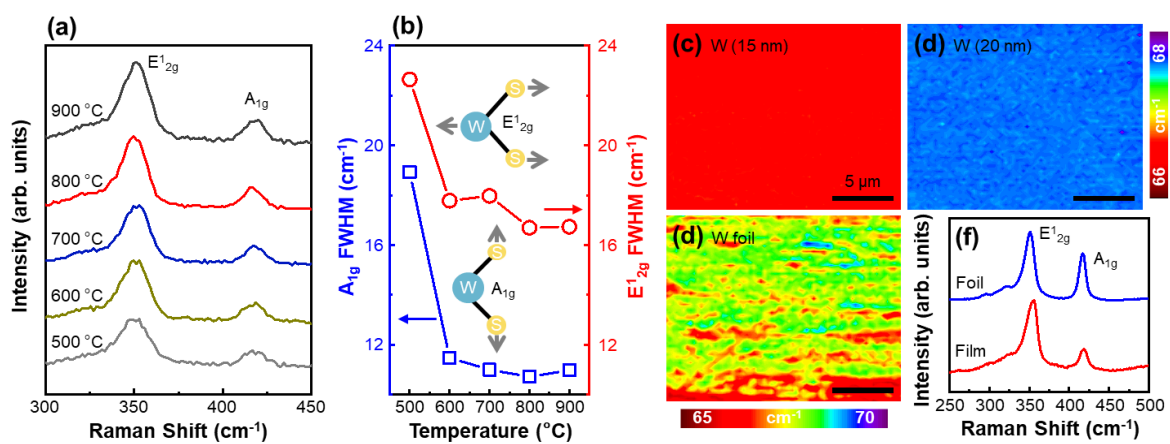

**Figure 2**. **(a)** Raman spectra of WS$_2$ thin films grown by T-CVD at various synthesis temperatures, highlighting the characteristic $A_{1g}$ and $E^1_{2g}$ vibrational modes. **(b)** Plots of the FWHM for the $A_{1g}$ (red circles, right axis) and $E^1_{2g}$ (blue squares, left axis) modes as a function of the WS$_2$ growth temperature. The data indicate a clear dependence of peak broadening on the formation temperature, reflecting changes in crystallinity and strain levels in the WS$_2$ layer. **(c)**, **(d)**, and **(e)** Raman mapping images of the $A_{1g}$ - $E^1_{2g}$ frequency difference for WS$_2$ films transferred onto quartz after T-CVD growth at 800 °C. In **(c)** and **(d)**, WS$_2$ was grown on W thin films of 15 nm and 20 nm thickness, respectively, while **(e)** shows the corresponding map for WS$_2$ grown on a W foil. The color scale indicates spatial variations in the $A_{1g}$ - $E^1_{2g}$ difference, providing insight into film uniformity and local strain or thickness variations. **(f)** Representative Raman spectra extracted from the mapping data, comparing WS$_2$ transferred from the W thin film (red) and the W foil (blue).

**Figure 3** provides complementary XPS evidence for the chemical evolution from tungsten oxide to WS$_2$ during sulfurization. **Figure 3(a)** shows wide-scan survey spectra collected before (bottom, red) and after (top, blue) sulfurization at 800 °C. Prior to sulfurization, the spectrum is dominated by W and O core levels (e.g., W 4d, W 4f, O 1s), indicating a tungsten oxide surface layer. After sulfurization, strong S 2s and S 2p peaks appear while the O 1s intensity is markedly reduced, confirming incorporation of sulfur and substantial removal of oxygen.

High-resolution W 4f spectra are presented in **Figure 3(b)**. Before sulfurization, the W $4f_{7/2}$ and W $4f_{5/2}$ peaks are located at binding energies characteristic of $WO_2$. After sulfurization, the W 4f doublet shifts to ~32.85 eV and ~35.0 eV, accompanied by an additional component at ~38.5 eV, consistent with W in a sulfide environment within $WS_2$. Similarly, the high-resolution S 2p spectrum in **Figure 3(c)** shows a well-resolved S $2p_{3/2}$ - $2p_{1/2}$ doublet at 162.4 eV and 163.6 eV, characteristic of $S_2-$ in $WS_2$. The reduction of O-related features together with the emergence of $WS_2$-like W 4f and S 2p lines demonstrates an efficient transformation from $WO_2$ to $WS_2$ under the chosen sulfurization conditions. These results are in good agreement with previous work on the sulfurization of tungsten suboxides to grow $WS_2$ thin films [28].

The sulfurization mechanism is schematically illustrated in **Figure 3(d)**. Initially, a $WO_2$ layer resides on the $SiO_2$ substrate. Upon heating to 800 °C in the presence of sulfur vapor, oxygen atoms are progressively replaced by sulfur atoms, leading to the formation of layered $WS_2$ with a hexagonal structure. This chemical conversion, corroborated by the XPS analysis, underpins the structural and optical properties observed in the Raman measurements.

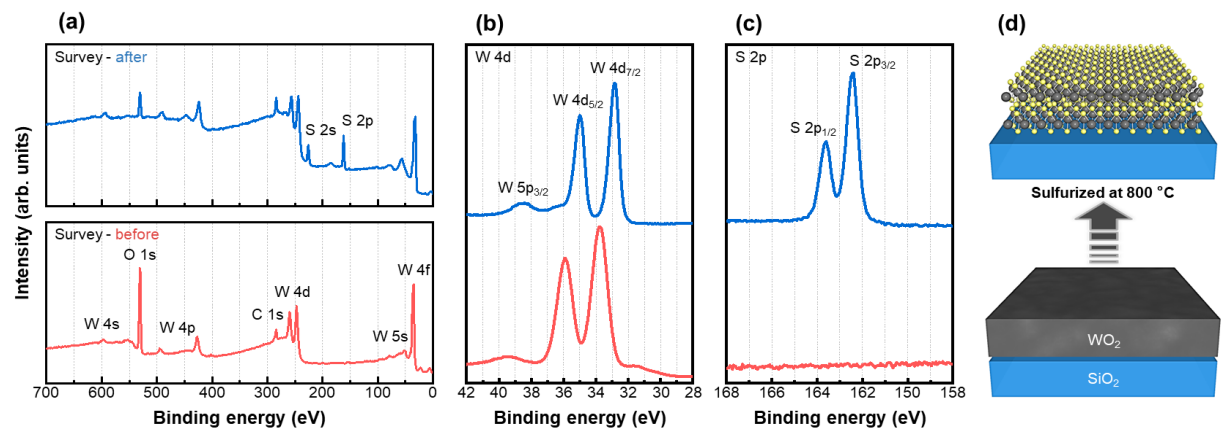

**Figure 3**. **(a)** Survey XPS spectra before (red) and after (blue) sulfurization at 800 °C. Initially, W and O peaks dominate, indicating a $WO_2$ layer, while subsequent sulfurization yields new S peaks and reduced O signals, confirming $WS_2$ formation. **(b)** and **(c)** High-resolution XPS spectra of the W 4f and S 2p regions, respectively. Prior to sulfurization, the W 4f spectrum corresponds to $WO_2$. Post-sulfurization, the W 4f peaks shift to about 32.85 eV, 35.0 eV, and 38.5 eV, and new S 2p peaks at roughly 162.4 eV and 163.6 eV emerge, indicative of $WS_2$. The reduced O intensity underscores the transition from oxide to sulfide. **(d)** Schematic illustration of the sulfurization process. $WO_2$ on the $SiO_2$ substrate is converted to layered $WS_2$ via O-S substitution during high-temperature sulfurization. The XPS results validate this transformation by showing distinct W 4f shifts, new S 2p signals, and diminished O 1s peaks, confirming the effective formation of $WS_2$.

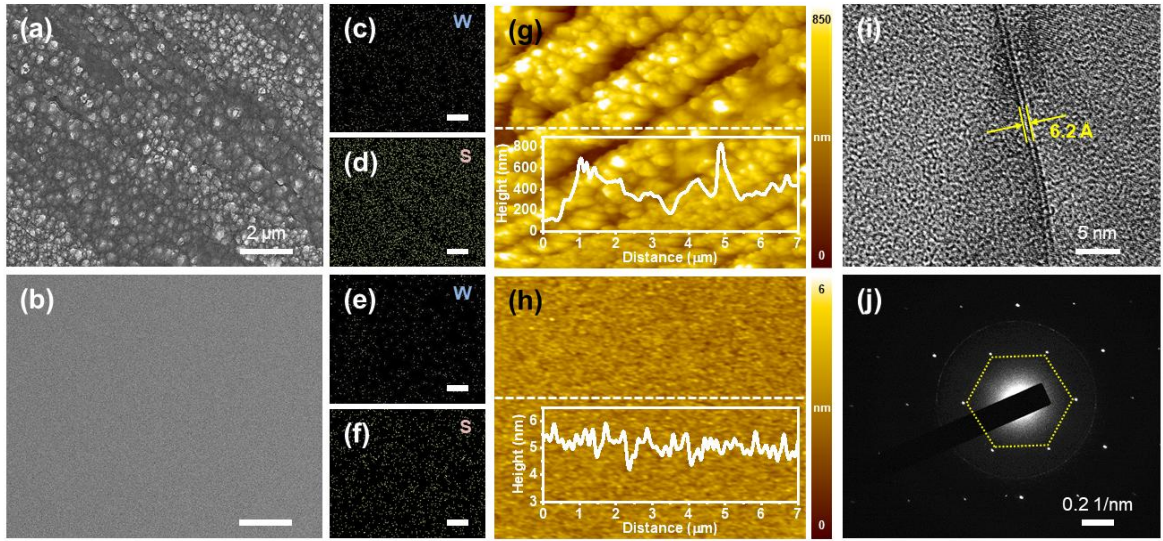

**Figure 4**. **(a)** SEM image of WS$_2$ synthesized from W foil and transferred to a SiO$_2$ substrate, showing island-like structures with significant surface roughness. **(b)** SEM image of WS$_2$ grown from a W thin film under otherwise identical conditions, demonstrating a notably more uniform and continuous morphology. EDS mapping in **(c)** and **(d)** corresponds to the W and S distributions, respectively, for the sample shown in **(a)**, highlighting that the island regions predominantly consist of tungsten disulfide. Similarly, **(e)** and **(f)** present EDS maps of W and S for the thin-film-derived sample in **(b)**, revealing a homogeneous composition across the surface. AFM characterization in **(g)** for the WS$_2$ grown from W foil confirms the presence of significant topographical variations, with an RMS roughness of ~390.62 nm; the inset line profile indicates steep height fluctuations within individual island features. In contrast, **(h)** shows an AFM image of the WS$_2$ grown from a W thin film, exhibiting a much lower RMS roughness of ~5.18 nm and a comparatively flat topology, as corroborated by the inset line profile. Finally, **(i)** and **(j)** provide TEM and SAED analyses of representative WS$_2$ domains, revealing a hexagonal lattice structure consistent with layered transition metal dichalcogenides and confirming that the synthesized films, despite marked differences in surface roughness, retain the expected crystalline phase.

The influence of the W precursor type on WS$_2$ morphology and crystallinity is examined in **Figure 4**. SEM images in **Figures 4(a)** and **4(b)** compare WS$_2$ films derived from W foil and sputtered W thin film, respectively. The foil-derived WS$_2$ (**Figure 4(a)**) exhibits a densely packed, island-like morphology with pronounced surface roughness and large grains. In contrast, the thin-film-derived WS$_2$ (**Figure 4(b)**) presents a smooth, featureless contrast consistent with a continuous, uniform film.

Elemental distributions obtained by energy-dispersive X-ray spectroscopy (EDS) are shown in **Figures 4(c) - (f)**. For the foil-derived film, the W (**Figure 4(c)**) and S (**Figure 4(d)**) maps both display clustered regions that follow the island morphology observed in SEM, indicating inhomogeneous WS$_2$ growth. In contrast, the W (**Figure 4(e)**) and S (**Figure 4(f)**) maps for the thin-film-derived WS$_2$ are spatially uniform, confirming that both elements are evenly distributed across the sample and supporting the conclusion of homogeneous film coverage.

Atomic force microscopy (AFM) further highlights the stark difference in surface topography (**Figures 4(g)** and **4(h)**). The WS$_2$ grown from W foil exhibits significant height variations approaching ~1 μm and an RMS roughness of ~390.62 nm, consistent with a three-dimensional island growth mode. The inset height profile in **Figure 4(g)** shows multiple peaks associated with these tall islands. In contrast, the thin-film-derived WS$_2$ has an RMS roughness of only ~5.18 nm, and the corresponding height profile in **Figure 4(h)** shows small fluctuations around a nearly flat baseline, characteristic of a conformal thin film. These observations suggest that the rough, polycrystalline surface and grain boundaries of the foil promote non-uniform nucleation and localized vertical growth, while the sputtered W film provides a smoother, more uniform template that favors lateral, layer-by-layer growth. Differences in local diffusion pathways for W and S species during T-CVD likely further amplify these morphological contrasts.

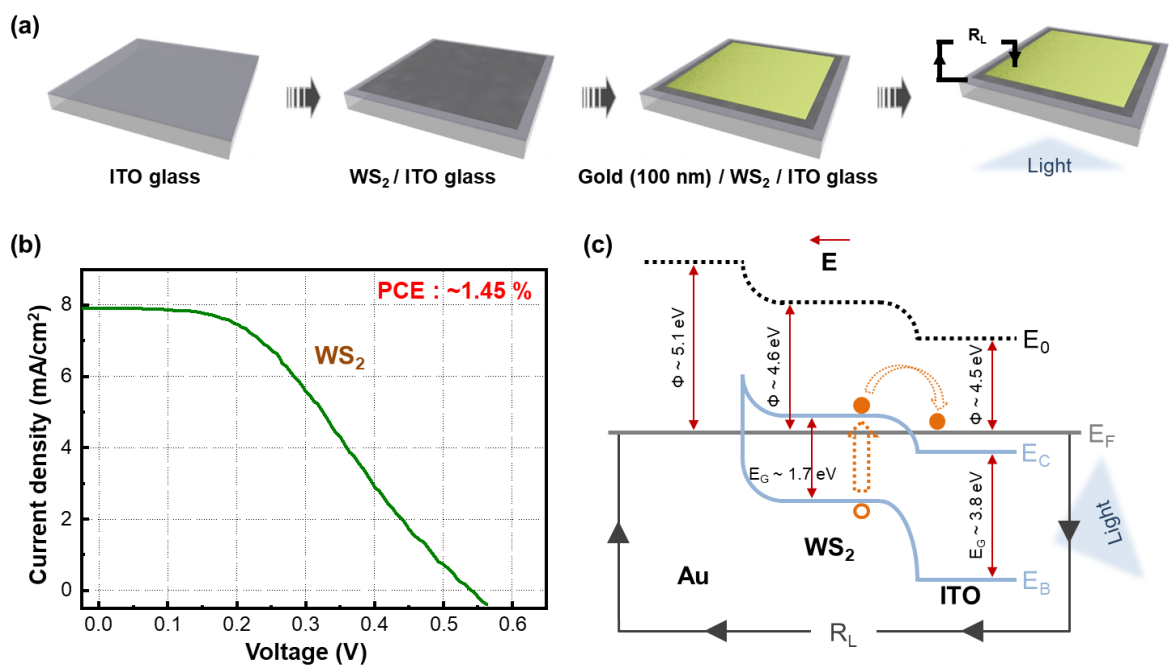

**Figure 5. (a)** Schematic illustration of the sequential fabrication and transfer process of WS$_2$ thin films onto ITO-coated glass using a PMMA-assisted wet-transfer method. Gold electrodes (~100 nm thick) were deposited on the WS$_2$ film to complete the Au/WS$_2$/ITO Schottky-barrier solar cell structure. **(b)** Current density-voltage (J-V) characteristics of the Au/WS$_2$/ITO solar cell measured under AM 1.5 illumination, exhibiting a short-circuit current density (J$_{SC}$) of approximately 7.91 mA cm$^{-2}$, an open-circuit voltage (V$_{OC}$) of ~495 mV, a fill factor (FF) of around 40 %, and a power conversion efficiency (PCE) of ~1.45 %. **(c)** Qualitative energy band diagram illustrating the Schottky barrier formed at the Au-WS$_2$ interface, charge-carrier generation by photon absorption, and subsequent carrier separation and transport processes. Defects in WS$_2$ layers may lead to increased electron-hole pair recombination and reduced device performance.

The crystallinity of the $WS_2$ films is confirmed by high-resolution TEM (HR-TEM) and selected-area electron diffraction (SAED), shown in **Figures 4(i)** and **4(j)**. The HR-TEM image (**Figure 4(i)**) reveals clear lattice fringes with an interlayer spacing of ~6.2 Å, in good agreement with the (002) basal-plane spacing of hexagonal $WS_2$. The SAED pattern in **Figure 4(j)** displays a set of sixfold-symmetric diffraction spots, characteristic of a hexagonal layered structure with good crystallinity. Together with the Raman and XPS analyses, these results demonstrate that the optimized sulfurization process produces crystalline $WS_2$ and that the sputtered W film precursor is particularly advantageous for achieving large-area films with low roughness and uniform thickness properties that are critical for optoelectronic device applications.

Finally, the photovoltaic performance of the WS2 thin films is evaluated using an $Au/WS_2/ITO$ Schottky-barrier solar cell architecture, shown in **Figure 5**. **Figure 5(a)** schematically illustrates the device stack and fabrication sequence. $WS_2$ grown on a sacrificial substrate is first coated with a PMMA support layer and then released by selectively etching the underlying $W/SiO_2$ stack. The resulting $PMMA/WS_2$ membrane is transferred onto ITO-coated glass, after which the PMMA is removed to form the $WS_2/ITO$ junction. Subsequently, ~100 nm-thick Au top contacts are deposited to form the Schottky interface. The completed device is then illuminated while the current-voltage characteristics are measured across an external load.

The measured J-V curve under illumination is shown in **Figure 5(b)**. The representative device exhibits a short-circuit current density ($J_{SC}$) of approximately 7.91 mA·cm$^{-2}$, an open-circuit voltage ($V_{OC}$) of ~0.495 V, and a fill factor (FF) of ~40 %, yielding a power conversion efficiency (PCE) of ~1.45 %. These values compare favorably with earlier $WS_2$-based photovoltaic devices that used mechanically exfoliated flakes or less uniform CVD films, which typically reported PCEs below 1 % due to their limited lateral dimensions and high defect densities, which impede efficient carrier transport [15,39,40,43]. The relatively high $J_{SC}$ achieved here indicates effective photon absorption and charge-carrier generation in the $WS_2$ film. At the same time, the moderate $V_{OC}$ and FF suggest that there is still room for improvement through further optimization of the metal contacts, interface engineering, and defect passivation [31,38,41,42].

A qualitative band diagram of the $Au/WS_2/ITO$ Schottky junction under illumination is depicted in **Figure 5(c)**. The difference between the Au work function and the $WS_2$ electron affinity establishes a Schottky barrier for electrons at the $Au/WS_2$ interface. In contrast, the $WS_2/ITO$ interface forms a relatively low barrier for hole collection. When illuminated, photons with energy greater than the $WS_2$ band gap (~1.7 eV) generate electron-hole pairs in the $WS_2$ layer. The built-in electric field across the junction separates these photogenerated carriers:

electrons drift toward the Au contact and holes toward the ITO, producing a photocurrent in the external circuit. Structural defects, interfacial states, or surface contamination in $WS_2$ can act as non-radiative recombination centers, reducing both $V_{OC}$ and $J_{SC}$ [41,42]. Conversely, the high crystallinity, smooth morphology, and clean interfaces achieved via optimized sulfurization and careful transfer help suppress recombination and enable the observed PCE of 1.45 %. These results underscore the potential of sputtered-W-derived $WS_2$ thin films as scalable active layers for Schottky-barrier solar cells and other large-area optoelectronic and energy-harvesting devices.

## 4. Conclusion

In summary, we have demonstrated a scalable route for synthesizing large-area $WS_2$ thin films by sulfurizing tungsten foils and sputtered W thin films in a custom-built thermal CVD system. By systematically varying the sulfurization temperature from 400 to 1000 °C, Raman spectroscopy identified 800 °C as the optimal growth condition, yielding the narrowest FWHM of the $E^1_{2g}$ and $A_{1g}$ modes and thus the highest crystalline quality. XPS analysis confirmed the chemical conversion from tungsten oxide to $WS_2$. At the same time, SEM, AFM, and TEM revealed that sputtered-W-derived $WS_2$ forms uniform, low-roughness films with well-defined layered crystallinity, in contrast to the thicker, island-like, and highly non-uniform morphology obtained from W foil precursors.

Using a PMMA-assisted wet-transfer process, these $WS_2$ thin films were integrated onto various substrates, including quartz and ITO-coated glass, without significant degradation of structural or optical quality. The transferred $WS_2$ was then employed as the photoactive layer in Au/$WS_2$/ITO Schottky-barrier solar cells, which exhibited a short-circuit current density of 7.91 mA·cm$^{-2}$, an open-circuit voltage of 0.495 V, a fill factor of ~40 %, and a power conversion efficiency of 1.45 %. This performance compares favorably with earlier $WS_2$-based photovoltaic devices that relied on exfoliated flakes or less optimized CVD films. It highlights the importance of film uniformity, crystallinity, and smooth interfaces for efficient carrier generation and extraction.

Although the PCE remains modest relative to commercial silicon photovoltaics, the combination of a simple T-CVD process, controllable film thickness, and wafer-scale uniformity establishes sputtered-W-derived $WS_2$ as a promising platform for large-area optoelectronic and energy-harvesting applications. Future improvements are expected from systematic optimization of metal contacts, interface engineering, controlled doping, and defect passivation, as well as from integrating WS$_2$ into van der Waals heterostructures and flexible device architectures. These advances could further enhance device performance and enable WS$_2$-based technologies in lightweight, mechanically flexible, and radiation-tolerant systems relevant to both terrestrial and space environments.